\documentclass[aps,floatfix,prstab,preprint,a4paper,%
superscriptaddress,showpacs]{revtex4}
\begin{document}

\title{Comment on  equivalence between quantum phase transition phenomena in
radiation-matter and magnetic systems }
\author{J. G. Brankov}
\email{brankov@bas.bg} \affiliation{Institute of Mechanics, acad G
Bonchev 4, 1113 Sofia, Bulgaria}

\author{N.S. Tonchev}
\email{tonchev@issp.bas.bg} \affiliation{Institute of Solid State
Physics, 72 Tzarigradsko Chauss\'ee, 1784 Sofia, Bulgaria}

\author{V. A. Zagrebnov}
\email{zagrebnov@cpt.univ-mrs.fr}
 \affiliation{Universit\'e de la
Mediterran\'ee (Aix-Marseille II) and Centre de Physique
Th\'eorique, Luminy-Case 907, 13288 Marseille, Cedex 09, France}

\begin{abstract}

In this Comment we show that the temperature-dependent effective
Hamiltonian derived by Reslen {\it et al} [Europhys. Lett., {\bf
69} (2005) 8] or that one by Liberti and Zaffino
[arXiv:cond-mat/0503742] for the Dicke model cannot be correct for
any  temperature. They both violate a rigorous result. The former
is correct only in the  quantum  (zero-temperature) limit while
the last one only in the classical (infinite temperature) limit.
The fact that the Dicke model belongs to the universality class of
the infinitely coordinated transverse-field XY model is known for
more then 30 years.
\end{abstract}


\maketitle

Recently a rapidly growing body of papers  point to a connection
between the thermodynamic and entanglement properties of the Dicke
model and the thermodynamic and entanglement properties of an
infinitely  coordinated, transverse-field XY model (see \cite{R05,
LZ05} and refs. therein). The attempt is to classify and
understand the entanglement properties of the Dicke Hamiltonian by
looking at an effective spin- spin exchange Hamiltonian.

For example in  \cite{R05} the authors show that instead of the
original Dicke Hamiltonian ($\hbar=c=\omega=1)$
\begin{equation}\label{D}
H_{Dicke}=a^{\dagger}a + \epsilon J_{z} -
\left[\frac{2\lambda}{N^{1/2}}\right](a^{\dagger} + a)J_{x}
\end{equation}
for studying the thermodynamic properties one can use the following
temperature-dependent effective Hamiltonian:
\begin{equation}\label{CT}
H_{qb}^{2}(\beta)=\epsilon J_{z} -
\left[\frac{2\lambda}{N^{1/2}}\right]^{2}\left[1+
\frac{2}{\beta(h(\beta)+1)}\right]J_{x}^{2},
\end{equation}
where $J_{z} = \frac{1}{2}\sum_{i=1}^{N}\sigma_{i,z}$, $J_{x} =
\frac{1}{2}\sum_{i=1}^{N}(\sigma^{\dagger}_i +\sigma_i)$,
and $h(\beta)=(e^{\beta}-1)^{-1}$
is the Bose factor which determines the average photon number in an
isolated cavity (single radiation mode of energy $\omega =1$) at inverse
temperature $\beta =(k_{\rm B}T)^{-1}$.

In some agreement with \cite{R05}, "to investigate the connection
between the Dicke and the collective one-dimensional Ising model",
in \cite{LZ05} the authors suggested the following effective
Hamiltonian (for the sake of convenience, here and below we use
the notation of \cite{R05}):
\begin{equation}
\label{LZ} H^{eff}_{A}(\beta)=\epsilon J_{z} -
\frac{\beta}{2}\left[\frac{2\lambda}{N^{1/2}}\right]^{2}\coth\left(\frac{\beta}{2}\right)J_{x}^{2}.
\end{equation}

The both statements are {\textit{wrong}}, because it has been
proven rigorously in \cite{BZT75} that Hamiltonian (\ref{D}) is
equivalent in the thermodynamic limit to the Hamiltonian (see Eqs.
(2) and (34) in \cite{BZT75} at $\mu =1$):
\begin{equation}\label{AH}
\left. H_{s}\right|_{\mu=1}=\epsilon J_{z} -
\left[\frac{2\lambda}{N^{1/2}}\right]^{2}J_{x}^{2}.
\end{equation}
This result was formalized as a rigorous mathematical statement for a much
larger class of models of matter interacting with boson fields, a particular
case of which is the Dicke model, see Theorem 4.1 in \cite{B}. For the free
energy densities
\begin{equation}
f_{N}[H_{Dicke}]= -\frac{1}{\beta N}\ln {\rm Tr} \exp(-\beta H_{Dicke}), \quad
f_{N}[\left. H_{s}\right|_{\mu=1}]= -\frac{1}{\beta N}\ln {\rm Tr}
\exp(-\beta \left. H_{s}\right|_{\mu=1}),
\end{equation}
we have obtained the following estimates:
\begin{equation}\label{est}
-\delta^{H}_{N}\leq f_{N}[H_{Dicke}] - f_{N}[\left. H_{s}
\right|_{\mu=1}]\leq \delta^{B}_{N},
\end{equation}
where $\delta^{H}_{N}=O(N^{-1/2})$ and $\delta^{B}_{N}=O(N^{-1}\ln
N)$ as $N\rightarrow \infty$. The result (\ref{est}) is
independent of the temperature and excludes the relation of
Hamiltonians (\ref{CT}) or (\ref{LZ}) with the thermodynamics of
the Dicke model (\ref{D}).
 About the same
time this statement was obtained by completely different methods,
see \cite{HL}-\cite{FSV}.

Note that at zero temperature Hamiltonian (\ref{CT}) coincides
with  Hamiltonian (\ref{AH}). Precisely this makes correct the
further calculations in the paper of Reslen {\it et al} \cite{R05}
as far as they are carried out at zero temperature. In this
connection, some doubts arise about the correct implementation of
the cumulant projection method suggested by Polatsek and Becker
\cite{PB97} for construction of "size-consistent" effective
Hamiltonians, at least with respect to the Dicke model. The above
cited authors claim that "our derivation is general, and can be
applied to any temperature, and to any sort of splitting of the
Hamiltonian" \cite{PB97}. Provided the Hamiltonian (\ref{CT}) is
correctly derived by that method in \cite{R05}, one faces a
counter-example of its applicability.

In the other limit $\beta \to 0$,  Liberti and Zaffino \cite{LZ05}
call it "classical" , Hamiltonian (\ref{LZ}) also coincides with
Hamiltonian (\ref{AH}). The source of incorrectness in obtaining
Hamiltonian (\ref{LZ}) seems to be more clear. In calculating the
partition function on the basis of the Zassenhaus formula it is
not sufficient to keep only the lowest-order terms (see Appendix B
in \cite{LZ05}). One can see that in the "quantum" limit $\beta
\to \infty$  Hamiltonian (\ref{LZ}) is not a {\textit bona fide}
Hamiltonian since it does not belong to the trace-class operators.

Furthermore, even disregarding the spurious temperature-dependent
term in (\ref{CT}) (or to consider only the classical limit in
(\ref{LZ})), the statement in \cite{R05} that "the physics of the
QPT (quantum phase transition) in the Dicke model is indeed
captured by the effective Hamiltonian of Eq. (\ref{CT})" may be
taken only on trust. The concept of thermodynamic equivalence
includes the equality of all the observable characteristics of the
equilibrium state, such as average values of local operators and
correlation functions of any finite order. Proving such a
statement is quite complicated problem and needs special
mathematical technics \cite{B}.

Our comments so far concerned thermodynamic properties only. The
revived interest on the Dicke model is caused by the perceived
relations between  thermodynamic  and entanglement properties.
 It is known that mean-field models, as the one under
consideration, cannot provide nontrivial entanglement properties
since the problem is effectively a single body one in the
thermodynamic limit. That is why one has to consider finite-N
systems \cite{V04} and the entanglement properties are necessarily
tested in the framework of the finite-size scaling (FSS) theory.
 May
be the instructive part of Ref. \cite{R05} treats the entanglement
properties. In this case the following question arises: If two
Hamiltonians generate equivalent (in some sense) critical
behaviour in the thermodynamic limit, are their finite-size
properties similar? The answer to that questions is: not always.
In the case under consideration the thermodynamic equivalence of
the models (\ref{D}) and (\ref{AH}) has been proved by the method
of the approximating Hamiltonian, see e.g. \cite{B,BDT}. The
application of this method is based on the fact that (\ref{D}) and
(\ref{AH}) have a common approximating Hamiltonian and the proof
of their thermodynamic equivalence passes through the limit of an
infinite system. For a finite $N$ we have just the lower and upper
bounds on the difference of the free energies per spin
(\ref{est}). The closeness of the finite-size properties of the
original and the effective model poses a subtle problem. It has
been shown that in some cases the FSS scaling functions for the
original and the approximating Hamiltonians are different
\cite{B90}. If this is so for thermodynamic functions, the problem
of closeness of the measures of entanglement, such as the
concurrence discussed in \cite{R05}, is still more problematic,
since it probes the internal structure of the ground-state in a
more detailed way. Therefore, it is not surprising that the
maximum value of the concurrence obtained in \cite{R05} was found
to be overestimated by the effective Hamiltonian as compared to
the original one (\ref{D}). Up to now the link between
entanglement and critical properties is not completely understood.
In fact one must be able to control the convergence of the quantum
Gibbs state for (\ref{D}) to the state corresponding to the
thermodynamically equivalent effective Hamiltonian (\ref{AH}). The
accuracy of this control by any of the known methods
\cite{BZT75}-\cite{FSV} needs supplementary investigations. That
is why we think the use of thermodynamically equivalent effective
Hamiltonians for studying entanglement properties is still an open
problem.

\end{document}